\documentclass[preprint,12pt]{elsarticle}
\usepackage{amssymb}
\usepackage{amsmath}
\usepackage{bm}
\usepackage{mathtools}
\usepackage[pdftex]{hyperref}
\usepackage[pdftex]{color}

\renewcommand{\d}{d}
\newcommand{\e}{e}
\newcommand{\imag}{i}

\newcommand{\phiq}{\varphi_{\bm{q}}}

\journal{Journal of Subatomic Particles and Cosmology}

\begin{document}
\begin{frontmatter}

\title{Higher partial waves in femtoscopy}
\author[a,b]{Koichi Murase}
\author[a,b]{Tetsuo Hyodo}
\affiliation[a]{organization={Department of Physics, Tokyo Metropolitan University},
            addressline={1-1 Minami-Osawa},
            city={Hachioji},
            postcode={192-0397},
            state={Tokyo},
            country={Japan}}
\affiliation[b]{ organization={RIKEN Interdisciplinary Theoretical and
Mathematical Science Program (iTHEMS)}, addressline={2-1 Hirosawa},
city={Wako}, postcode={351-0198}, state={Saitama}, country={Japan} }

\begin{abstract}
Femtoscopy is recently gaining more attention as a new approach complementary
to scattering experiments for constraining hadron-hadron interactions.  We
discuss the effect of higher partial waves on the two-particle correlation
function, which has been neglected in traditional formulae used in the
femtoscopy analyses.  We consider the partial-wave expansion of the scattering wave
function in the Koonin--Pratt formula to give the correction to the correlation
function by a sum of the contributions from each partial wave.  We also
generalize the Lednick\'y--Lyuboshitz formula, which was originally derived for
the $s$-wave interaction, to include higher partial waves.  We find a compact
representation of the generalized Lednick\'y--Lyuboshitz formula given by the
backward scattering amplitude $f(\theta = \pi)$ and the Fourier--Laplace
transform of the relative source function, which gives an insight into the
structure of the Lednick\'y--Lyuboshitz formula and its relation to the optical
theorem.  We numerically demonstrate the significance of higher partial waves
with resonances using the square potential well.  Also, the generalized
Lednick\'y--Lyuboshitz formula turned out to be broken for higher partial
waves, which indicates the importance of the centrifugal force with the higher
partial waves.
\end{abstract}




\end{frontmatter}


\section{Introduction}

Recently, femtoscopy in
high-energy nuclear collisions has been gathering attention as a new approach
to hadron-hadron interaction~\cite{Ohnishi:1998at, Morita:2014kza, ExHIC:2017smd, Fabbietti:2020bfg}.  Femtoscopy allows us to access information
about the particle source size and the wave function through the two-particle
correlation function, which is actively measured in experiments
recently~\cite{STAR:2014dcy, STAR:2015kha, ALICE:2018ysd, ALICE:2019hdt,
ALICE:2019buq, ALICE:2019gcn, ALICE:2019eol, ALICE:2020mfd, ALICE:2021njx,
ALICE:2021cyj, ALICE:2021cpv, ALICE:2021szj, ALICE:2022uso, ALICE:2022enj,
ALICE:2022yyh, ALICE:2023wjz, ALICE:2024bhk}.
The two-particle correlation function for a pair of particle species
1 and 2 is defined as
\begin{align}
  C_{12}(\bm{p}_1, \bm{p}_2) = C_{12}(\bm{q}, \bm{P})
  &= \dfrac{E_1 E_2\dfrac{\d{N_{12}}}{\d\bm{p}_1\d\bm{p}_2}}{E_1\dfrac{\d{N_1}}{\d\bm{p}_1} E_2\dfrac{\d{N_2}}{\d{\bm{p}_2}}},
\end{align}
where $\bm{p}_1$ and $\bm{p}_2$ are the momenta of the particles, and $E_1 =
E_1(\bm{p}_1)$ and $E_2 = E_2(\bm{p}_2)$ are the corresponding energies.  The
momentum spectra of the particles are given by $\d{N_1}/\d{\bm{p}_1}$ and
$\d{N_2}/\d{\bm{p}_2}$.  The pair distribution
$\d{N_{12}}/\d{\bm{p}_1}\d{\bm{p}_2}$ denotes the probability density that we
find particles $1$ and $2$, with their respective momenta being $\bm{p}_1$ and
$\bm{p}_2$, simultaneously in a single event. The correlation function $C_{12}$ is
also written as a function of the relative and total momenta of the
two-particle system, $\bm{q}$ and $\bm{P}$, instead of $\bm{p}_1$ and
$\bm{p}_2$.  In practice, the correlation function can be averaged for $\bm{P}$
with a specific frame choice, such as the pairwise rest frame, to define $C_{12}(\bm{q})$.

A perfect description of the correlation function in the high-energy nuclear
collisions would require a theoretical understanding of dynamical particle
production in the complicated processes of the collision reactions.  Instead,
based on a number of assumptions, the Koonin--Pratt (KP)~\cite{Koonin:1977fh,
Pratt:1984su, Bauer:1992ffu} and Lednick\'y--Lyuboshitz (LL)~\cite{Lednicky:1981su} formulae
give useful expressions for the two-particle correlation function in
terms of the source function, $S(\bm{r})$, and the wave function,
$\phiq^{(-)}(\bm{r})$, with $\bm{r}$ being the relative coordinates of the
emission points of the two particles.  Those formulae have proven their
convenience and power in analyzing the correlation function in existing
femtoscopy studies.  However, as the experimental data for various particle
pairs are becoming available with better statistics, precise and systematic
determination of the hadronic interactions based on a realistic setup is
gaining more importance.  Some of the assumptions so far we relied upon may not
be compatible with a realistic setup. It is important to revisit each
assumption and consider its role.

In this study, we focus on the assumption of the $s$-wave interaction (i.e.,
the interaction for the orbital angular momentum $l = 0$) used in the LL
formula, where the contributions from $p$- and $d$-waves are neglected for
simplicity. In general, the correlation function includes contributions from
all partial waves.  The effect of higher partial waves is expected to be
particularly significant in the presence of resonances in higher partial
waves. For example, a peak around $|\bm{q}|\simeq 240~\text{MeV}$ observed in
the $K^-$-$p$ correlation~\cite{ALICE:2019gcn} is attributed to the $d$-wave
resonance $\Lambda(1520)$.  This peak in the correlation function has been
attempted to be fit by a Breit--Wigner function~\cite{Kamiya:2019uiw} on top of
the $s$-wave contribution, but its underlying justification and validity are
not obvious.  Recent ALICE results for the $\Lambda$-$K^-$
correlation~\cite{ALICE:2023wjz} also show peaks associated with higher partial
waves, such as the $\Omega$ peak via weak decay at $|\bm{q}|\simeq210.7~\text{MeV}$ in the $p$ wave
and the $\Xi(1820)$ peak at $|\bm{q}|\simeq398.9~\text{MeV}$ in the $d$ wave.
The importance of higher partial waves in the correlation function is recently
reported also in the studies of the three-body and
two-body calculations for the
$p$-$d$ correlation~\cite{Viviani:2023kxw, Rzesa:2024oqp, Torres-Rincon:2024znb}.
In this study, we try to extend the KP and LL formulae to include the
contributions from higher partial waves and examine their properties.  We
assume $\hbar=c=1$ and $\hbar c = 197.32~\text{MeV~fm}$ throughout this paper.

\section{Koonin--Pratt and Lednick\'y--Lyuboshitz formulae}

We assume the KP formula~\cite{Koonin:1977fh, Pratt:1984su} as a starting
point of this study and also consider the extension of the LL
formula~\cite{Lednicky:1981su}.  In this section, we overview the KP and LL
formulae, which are widely used in femtoscopy to analyze the $s$-wave
interaction.

The KP formula is based on a picture in which the two particles at the
positions $\bm{r}_1$ and $\bm{r}_2$ are isolated from the rest of the system at
an emission time and interact with each other as a two-body isolated system
until the two-particle distance becomes sufficiently large.  With an additional
approximation\footnote{This is typically justified by using the on-shell
approximation, where the particles are assumed to be emitted in on-shell
states, or the smoothness approximation, where the source function evaluated at
$\bm{q}_\mathrm{emit}$ is assumed to be close to the one at $\bm{q}$.}  that the relative
momentum $\bm{q}_\mathrm{emit}$ at emission can be replaced with the final momentum
$\bm{q}$, the KP formula gives the two-particle correlation function with the
integral:
\begin{align}
  C(\bm{q}) &= \int \d^3{\bm{r}}\, S_{12}(\bm{r}, \bm{q}) |\phiq^{(-)}(\bm{r})|^2,
  \label{eq:kpx.kp}
\end{align}
where $\bm{r} = \bm{r}_1 - \bm{r}_2$ is the relative coordinates of the
emission points $\bm{r}_1$ and $\bm{r}_2$.

The relative source function $S_{12}(\bm{r}, \bm{q})$, with the normalization
$\int_0^\infty \d^3{\bm{r}}\, S_{12}(\bm{r}, \bm{q}) = 1$, gives the distribution of the
relative positions $\bm{r}$ of the particle pairs with the relative momentum
$\bm{q}$.  The dependence of $S_{12}(\bm{r}, \bm{q})$ on $\bm{q}$ is usually
assumed to be negligible, and the relative source function is simply denoted as
$S(\bm{r}) := S_{12}(\bm{r}, \bm{q})$.

The relative wave function $\phiq^{(-)}(\bm{r})$ represents the amplitude of
finding the relative position of the two emission points at $\bm{r}$ when the
final relative momentum of the two particles is known to be $\bm{q}$.  The wave
function $\phiq^{(-)}(\bm{r})$ can be obtained by solving the Schr\"odinger
equation under the boundary condition matching the outgoing wave with the plane
wave of the asymptotic momentum $\bm{q}$ at $|\bm{r}|\to\infty$.  This is the
opposite of the standard two-body scattering problem, where the
incoming wave is normalized as a plane wave.  The solution is related to the
scattering wave function in the two-body scattering problem by the complex
conjugate: $\phiq^{(-)}(\bm{r}) = [\varphi_{\bm{q}}(-\bm{r})]^*$.  In the
following discussion, we consider the scattering wave function $\phiq(\bm{r})$
instead of $\phiq^{(-)}(\bm{r})$ to utilize the standard language in
the scattering theory; the wave function in Eq.~\eqref{eq:kpx.kp} may be simply
replaced with the scattering wave function because of the absolute value.  We
consider the non-relativistic case and assume a short-range local potential in
the Schr\"odinger equation:
\begin{align}
  \biggl[ -\frac1{2\mu} \nabla^2 + V(\bm{r}) \biggr] \phiq(\bm{r}) &= E_{\bm{q}} \phiq(\bm{r}),
  \label{eq:kpx.schro1}
\end{align}
where $\mu$ is the reduced mass of the two-body system, and the energy is
specified by the asymptotic momentum: $E_{\bm{q}} = q^2/(2\mu)$. The boundary
condition is given by
\begin{align}
  \phiq(\bm{r}) = \e^{\imag qz} + \frac{f(q, \theta)}r \e^{\imag qr} + \mathcal{O}\Bigl(\frac1{r^2}\Bigr), \quad (r\to\infty),
\end{align}
where the $z$-axis is taken to be the direction of the relative momentum
$\bm{q}$, and the polar angle $\theta$ is the angle between the vectors,
$\bm{r}$ and $\bm{q}$.  The function $f(q, \theta)$ is an undetermined function of
$q$ and $\theta$.  We here do not consider the symmetrization and antisymmetrization
associated with the identical two bosons and fermions; the following discussion
is only applicable to the non-identical particle pairs.

In the analyses for the $s$-wave interaction, we assume that only the $s$-wave
part of the wave function is modified by the interaction, and the rest is kept
to be that of the plane wave.  In this case, the wave function is written as
\begin{align}
  \phiq(\bm{r}) = \e^{\imag qz} - j_0(qr) + R_0(r),
  \label{eq:kp.wave.smod}
\end{align}
where $j_0(qr)$ is the spherical Bessel function of the first kind, and
$R_0(r)$ is the radial wave function of the $s$ wave\footnote{See
Eq.~\eqref{eq:kpx.wavex} for an explicit definition of $R_0(r)$.}. The first two
terms represent
the plane wave without the $s$-wave component.  If the source function is
spherical, $S(\bm{r}) = S(|\bm{r}|)$, applying Eq.~\eqref{eq:kp.wave.smod} to
Eq.~\eqref{eq:kpx.kp}, we obtain an expression of the correlation
function~\cite{Morita:2014kza} as
\begin{align}
  C(q) &= 1 + \int_0^\infty \d{r}\, 4\pi r^2 S(r) \bigl[|R_0(r)|^2 - |j_0(qr)|^2\bigr].
  \label{eq:kp.spherical}
\end{align}
This expression shows that the deviation of the correlation function from unity
is interpreted as the correction of the $s$-save change from $|j_0(qr)|^2$ to
$|R_0(r)|^2$.

The Lednick\'y--Lyuboshitz (LL) formula is another useful expression giving an
explicit form of the correlation function, which is based on the KP formula
with additional assumptions.  In the LL formula, the $s$-wave function in the
entire $r$ region is replaced with the asymptotic form at $r\to \infty$,
\begin{align}
  R_0^{\mathrm{asy}}(r) = \frac{\e^{\imag\delta_0(q)}}{qr}\sin[qr + \delta_0(q)],
  \label{eq:llx.llasy}
\end{align}
determined by the phase shift $\delta_0(q)$.  With the additional approximation
that the source function is written in the spherical Gaussian form,
\begin{align}
  S(r) &= \frac1{(4\pi R^2)^{3/2}}\exp\Bigl(-\frac{r^2}{4R^2}\Bigr),
  \label{eq:ll.gaussian-source}
\end{align}
the correlation function is given by
\begin{align}
  C_\mathrm{LL}(q) &= 1 + \frac{F_3(r_\mathrm{eff} / R)}{2R^2}|f_0(q)|^2 + \frac{2F_1(2qR)}{\sqrt{\pi}R}\Re f_0(q) -\frac{F_2(2qR)}{R} \Im f_0(q),
  \label{eq:ll.traditional}
\end{align}
where $F_1(x) = \int_0^x \d{t}\exp(t^2 - x^2)/x$, $F_2(x) = (1-\e^{-x^2})/x$,
and $F_3(x) = 1 - x/(2\sqrt{\pi})$.  The $s$-wave amplitude $f_0$\footnote{See
Eq.~\eqref{eq:llx.amplitude-expan} for an explicit definition of $f_0 =
f_0(q)$.} is related to the phase shift as $f_0(q) = (\e^{2\imag\delta_0(q)} -
1)/(2iq)$.  The second term of
$F_3(r_\mathrm{eff} / R)$ is called the effective range correction and
is introduced to partially compensate the approximation error by
$R^\mathrm{asy}_0(r)$:
\begin{align}
  \int \d{r}\, 4\pi r^2 S(r) \bigl[|R_0(r)|^2 - |R_0^{\mathrm{asy}}(r)|^2\bigr]
    &\simeq - \frac{r_\mathrm{eff}}{4\sqrt{\pi}R^3}|f_0(q)|^2,
\end{align}
where $r_\mathrm{eff} = (d/dq)^2 f_0(q)^{-1}|_{q=0}$ is the effective range.

\section{Higher partial waves}

In this section, we extend the existing formulae for the correlation function
with the $s$-wave correction to the general case with all partial waves.  We
first consider the generalization of the spherical-source KP
formula~\eqref{eq:kp.spherical} and the LL formula~\eqref{eq:ll.traditional} in
Secs.~\ref{sec:kpx} and \ref{sec:llx}, respectively. We also discuss the
structure of the LL formula in Sec.~\ref{sec:llopt}.

\subsection{KP formula with spherical source and higher partial waves}
\label{sec:kpx}

To extend the analysis for the general case where all the partial waves
contribute, we may fully expand the wave function $\phiq(\bm{r})$ in partial
waves:
\begin{align}
  \phiq(r, \theta) &= \sum_{l=0}^\infty (2l+1) \imag^l R_l(r) P_l(\cos\theta),
  \label{eq:kpx.wavex}
\end{align}
where $P_l(\cos\theta) \propto Y_l{}^0(\theta, \phi)$ are the Legendre
polynomials.  It is noted that $\phiq$ is independent of the azimuthal angle
$\phi$ due to the symmetry in the potential and the boundary condition, so
$\phiq(\bm{r})$ does not have the components $Y_l{}^m(\theta, \phi)$ ($m\neq
0$).  A set of the radial wave functions $\{R_l(r)\}_l$ for partial waves specifies
a total wave function.  For example, a plane wave $\phiq = \e^{\imag qz}$ is
specified by the spherical Bessel functions of the first kind, $R_l(r) =
j_l(qr)$, as is known in the Rayleigh expansion.  In the general case, the wave
function is obtained as the solution to the Schr\"odinger equation.  In the
case of the central force $V(r) = V(|\bm{r}_1 - \bm{r}_2|)$, the partial wave
function for each $l$ can be solved independently:
\begin{align}
  \biggl[
    -\frac{\partial^2}{\partial r^2} - \frac2{r}\frac{\partial}{\partial r}
    + \frac{l(l+1)}{r^2}
    + 2\mu V(r)
  \biggr] R_l(r) &= q^2 R_l(r).
  \label{eq:kpx.schro}
\end{align}

With the spherical source $S(r)$, we can apply Eq.~\eqref{eq:kpx.wavex} to
Eq.~\eqref{eq:kpx.kp} to obtain the correlation function in terms of $S(r)$ and
$R_l(r)$:
\begin{align}
  C(q)
  &= \sum_{l,l'=0}^\infty (2l+1)(2l'+1) \int d^3\bm{r}\, S(r)\, \imag^{l'-l} R_l^*(r)R_{l'}(r) P_l(\cos\theta)P_{l'}(\cos\theta)\notag \\
  &= \sum_{l=0}^\infty (2l+1) \int \d{r}\, 4\pi r^2 S(r) |R_l(r)|^2,
  \label{eq:kpx.kpx}
\end{align}
where we used the orthogonality of the Legendre polynomials to eliminate one
summation for the angular momentum $l'$.  We note that when the source function
is not spherical, all the pairs of the partial waves of different $l\neq l'$
interfere, and the correlation function cannot be written by a simple
summation for $l$.
Equation~\eqref{eq:kpx.kpx} gives the following relation for the plane wave,
$\phiq(\bm{r}) = \e^{\imag qz}$:
\begin{align}
  1 &= \sum_{l=0}^\infty (2l+1) \int \d{r}\, 4\pi r^2 S(r) |j_l(qr)|^2,
  \label{eq:kpx.kpx.plain}
\end{align}
where the unity on the left-hand side is directly obtained by the KP
formula~\eqref{eq:kpx.kp}.

Subtracting Eq.~\eqref{eq:kpx.kpx.plain} from Eq.~\eqref{eq:kpx.kpx}, we obtain
the correlation function in terms of the corrections:
\begin{align}
  C(q) &= 1 + \sum_{l=0}^\infty \Delta C_l(q), \label{eq:kpx.kpxd}
\end{align}
where $\Delta C_l(q)$ is the correction by the interaction in the $l$-th
partial wave:
\begin{align}
  \Delta C_l(q)
  &= (2l+1) \int \d{r}\, 4\pi r^2 S(r) \bigl[|R_l(r)|^2 - |j_l(qr)|^2\bigr]. \label{eq:kpx.kpxd.delta}
\end{align}
In the conventional case where only the $s$ wave is assumed to be modified, all
the higher-partial-wave contributions $\Delta C_l(q)$ ($l\ge1$) vanish so that
Eqs.~\eqref{eq:kpx.kpxd} and~\eqref{eq:kpx.kpxd.delta} reduce to the expression~\eqref{eq:kp.spherical}
known for the $s$-wave case.

\subsection{LL formula generalized for higher partial waves}
\label{sec:llx}

In addition to the spherical source assumption, the traditional LL formula
assumes the asymptotic form of the $s$-wave function for the entire spatial
region.
To extend the LL formula for higher partial waves, it is useful to first remind
us of the two levels of the asymptotic form of the wave function.  For a
short-range interaction, one can roughly
separate the space into three parts depending on the relative distance:
Region I is the spatial region within the interaction range $r <
r_\mathrm{int}$ where both the potential interaction and the centrifugal force
play roles.  Region II is the intermediate region where the potential is
negligible but the centrifugal force is still effective.  Region III is the
region where both the potential interaction and the centrifugal force are
negligible when considering the wave-function form.  The wave function in each region is given as the solution to the
Schr\"odinger equation with the corresponding approximation.
In Region I, the full Schr\"odinger equation~\eqref{eq:kpx.schro} needs to be
solved. The solution there depends on the explicit form of the potential
$V(r)$.  The asymptotic form of the wave function in Region II is obtained by
dropping the potential term $V(r)$ and solving Eq.~\eqref{eq:kpx.schro} with
only the first three terms on the left-hand side.  The solution is written by a
linear combination of the spherical Hankel functions $h_l(qr)$ and $h_l^*(qr)$:
\begin{align}
  R_l(r) \approx R_l^\mathrm{II}(r)
  &= \Bigl[\frac12 + iqf_l(q) \Bigr] h_l(qr) + \frac12 h_l^*(qr),
\end{align}
where $f_l(q)$ is the partial-wave amplitude defined by the expansion of the
scattering amplitude $f(q, \theta)$:
\begin{align}
  f(q, \theta) &= \sum_{l=0}^\infty (2l+1) P_l(\cos\theta) f_l(q).
  \label{eq:llx.amplitude-expan}
\end{align}
In this region, the potential information is rendered in the partial amplitudes
$f_l(q) = (\e^{2\imag\delta_l(q)} - 1)/(2iq)$ written by the phase shifts
$\delta_l(q)$, which are determined by the boundary condition with Region I\@.
In Region III, the asymptotic wave function is solved with the first two
terms on the left-hand side of Eq.~\eqref{eq:kpx.schro}, and the solution takes
the form:
\begin{align}
  R_l(r) \approx R_l^\mathrm{III}(r)
    &= \frac1{\imag^l} \frac{1}{\imag qr} \biggl\{\biggl[\frac12 +\imag qf_l(q)\biggr] \e^{\imag qr} - \frac12(-1)^l \e^{-\imag qr}\biggr\}.
  \label{eq:llx.asymp3}
\end{align}

For $l = 0$, the wave functions in Regions
II and III are the same $R_0^\mathrm{II}(r) = R_0^\mathrm{III}(r)$ because the
centrifugal potential $l(l+1)/r^2$ vanishes for $l=0$.  A natural extension of the
assumption of the asymptotic wave function~\eqref{eq:llx.llasy} to an arbitrary
angular momentum would be to use the asymptotic form $R_l^\mathrm{II}(r)$ or
$R_l^\mathrm{III}(r)$.  However, approximation of the wave function by
$R_l^\mathrm{II}(r)$ causes divergence of the correlation function
with $\delta_l \ne 0$ and $S(0) \ne 0$ because the asymptotic wave
function $R_l^\mathrm{II}(r)$ behaves as $\sim 1/r^{l+1}$ near the
origin.

Instead, we here consider $R_l^\mathrm{III}(r)$ for the wave function.
Substituting Eq.~\eqref{eq:llx.asymp3} for Eq.~\eqref{eq:kpx.kpx}, we obtain
\begin{align}
  C(q)
  &= \frac{4\pi}{q^2} \int \d{r}\,S(r) \sum_{l=0}^\infty (2l+1) \biggl\{1 - (-1)^{l+1} \Re\Bigl[\Bigl(\frac12+\imag qf_l\Bigr) \e^{2\imag qr}\Bigr]\biggr\}.
  \label{eq:llx.sub1}
\end{align}
In obtaining the
above formula, the probability conservation\footnote{Each term in the summation
of Eq.~\eqref{eq:prob-cons} vanishes if the orbital angular momentum $l$ is
conserved, but it is unnecessary because we only use the total probability
conservation here.} was used:
\begin{align}
  0 = \frac{\mu}q \int j_r(r, \theta, \phi) r^2 \d\Omega
    = \frac{4\pi}{q^2}\sum_{l=0}^\infty (2l+1) \biggl[\Bigl|\frac12 +\imag qf_l(q)\Bigr|^2 - \Bigl|\frac12(-1)^l\Bigr|^2\biggr],
  \label{eq:prob-cons}
\end{align}
where $\d\Omega = \d\phi\,\d\!\cos\theta$, and $j_r = (1/\mu)\Im(\phiq^*
\partial_r \phiq)$ is the $r$-component of the probability current with $\phiq$
given by $R_l^\mathrm{III}(r)$ through Eq.~\eqref{eq:kpx.wavex}.  The $r$
dependence of the asymptotic form cancel with the Jacobian $r^2$ on the
right-hand side.  In the
plain-wave case, the left-hand side of Eq.~\eqref{eq:llx.sub1} becomes unity,
while all $f_l$ on the right-hand side vanish.  Therefore, by subtracting the
plain-wave case of Eq.~\eqref{eq:llx.sub1} from itself, the correction to the
correlation function is given by the terms proportional to $f_l$:
\begin{align}
  C(q) &= 1+ \frac{4\pi}q\sum_{l=0}^\infty (2l+1)(-1)^l
    \Im [f_l(q) \hat S(-2\imag q)] \notag \\
  &= 1+ \frac{4\pi}q\sum_{l=0}^\infty (2l+1)(-1)^l
    [\Im f_l(q)\, \Re \hat S(-2\imag q) + \Re f_l(q)\, \Im \hat S(-2\imag q)],
    \label{eq:llx}
\end{align}
where $\hat S(-2\imag q) = \int_0^{\infty} \d{r}\, S(r) \e^{2\imag qr}$ is the
Fourier--Laplace transform of the radial source function.  In the case of the
Gaussian relative source~\eqref{eq:ll.gaussian-source}, the Fourier--Laplace
transform is given by
\begin{align}
  \hat S(-2\imag q) = \frac{\e^{-(2qR)^2}}{(4\pi)^{3/2} R^2} \biggl(\sqrt{\pi} + 2\imag\int_0^{2qR} dt \e^{t^2}\biggr),
  \label{eq:gaussian-FourierLaplace}
\end{align}
though Eq.~\eqref{eq:llx} holds for any spherical source function.  This gives
a naive generalization of the LL formula with higher partial waves.

Let us confirm that Eq.~\eqref{eq:llx} reproduces the original LL formula for
$s$ wave~\eqref{eq:ll.traditional} except for the effective range correction.
In the $s$-wave case, only
the correction from $l=0$ remains because $f_l(q) = 0$ for $l \ge 1$.  The
original LL formula~\eqref{eq:ll.traditional} has three correction terms
proportional to $|f_0(q)|^2$, $\Re f_0(q)$, and $\Im f_0(q)$ unlike the
generalized version~\eqref{eq:llx}, which has only two terms proportional to
$\Re f_l(q)$ and $\Im f_l(q)$.  However, we notice that the first and third
terms in the original LL formula can be combined using the optical theorem for
the partial amplitude, $|f_0(q)|^2 = (1/q) \Im f_0(q)$, so a simplified
expression of the original LL formula is found:
\begin{align}
  C_\mathrm{LL}(q)
  &= 1 + \frac{F_4(2qR, r_\mathrm{eff}/R)}{2qR^2}\Im f_0(q)
  +  \frac{2F_1(2qR)}{\sqrt{\pi}R}\Re f_0(q),
  \label{eq:llx.effective-range-correction.simplified}
\end{align}
where $F_4(x, y) = \e^{-x^2} - y/(2\sqrt{\pi})$.  The second argument of $F_4(x,
y)$ represents the effective range correction.  It is easy to show that the
coefficient functions in
Eq.~\eqref{eq:llx.effective-range-correction.simplified}, excluding the
effective range correction, match the real and imaginary parts of the
Fourier--Laplace transform of the relative Gaussian
source~\eqref{eq:gaussian-FourierLaplace}:
\begin{align}
  \frac{F_4(2qR, 0)}{2qR^2} = \frac{4\pi}q \Re \hat S(-2\imag q), \quad
  \frac{2F_1(2qR)}{\sqrt{\pi}R} = \frac{4\pi}q \Im \hat S(-2\imag q).
\end{align}

\subsection{Generalized LL formula and optical theorem}
\label{sec:llopt}

Noticing $(-1)^l = P_l(-1) = P_l(\cos\pi)$ and recalling
Eq.~\eqref{eq:llx.amplitude-expan}, we may further simplify Eq.~\eqref{eq:llx}
by taking the sum over $l$:
\begin{align}
  C(q) &= 1+ \frac{4\pi}q \Im [f(q, \theta = \pi) \hat S(-2\imag q)],
  \label{eq:llopt.llg}
\end{align}
where $f(q, \theta = \pi)$ is the backward scattering amplitude.  This
functional form is comparable to the optical theorem:
\begin{align}
  \sigma_\mathrm{tot} &= \frac{4\pi}q \Im f(q, \theta = 0),
  \label{eq:llopt.optical}
\end{align}
where $\sigma_\mathrm{tot}$ is the total cross section, and $f(q, \theta = 0)$
is the forward scattering amplitude.

The forward and backward amplitudes appearing in the optical theorem and the
correlation function are related to the structure of the asymptotic form of the
plane-wave expansion at the level of $R_l^\mathrm{III}(r)$:
\begin{align}
  \e^{\imag qz} &= \frac{2}{\imag qr}[\delta(1 - \cos\theta) \e^{\imag qr} - \delta(1 + \cos\theta) \e^{-\imag qr}] + \mathcal{O}\Bigl(\frac1{r^2}\Bigr).
  \label{eq:llopt.asymp-plane-wave}
\end{align}
A convention of the Dirac delta function $\delta(z)$ was chosen so that
$\int_0^\infty \d{z}\, \delta(z) g(z) = g(0)/2$ with $g(z)$ being an arbitrary
function analytic at $z=0$.  The above expression can be obtained by retaining
only the $\mathcal{O}(1/r)$ contributions of $R_l(r) = j_l(qr) \approx
(1/qr)\sin(qr - l\pi / 2)$ in the Rayleigh expansion~\eqref{eq:kpx.wavex} and
comparing them with the Legendre expansion of the delta functions.  This
asymptotic expression of the plane wave contains the two delta functions
peaking at $\cos\theta = \pm1$ (i.e., $\theta = 0, \pi$), which suggests that
the significance of forward and backward directions is already built in the
structure of the plane wave.

In fact, both the optical theorem~\eqref{eq:llopt.optical} and the generalized
LL formula~\eqref{eq:llopt.llg} can be easily derived by using
Eq.~\eqref{eq:llopt.asymp-plane-wave}.  The scattering wave function can be
written as
\begin{align}
  \phiq(\bm{r})
  &= \e^{\imag qz} + \frac{f(q, \theta)}r \e^{\imag qr} = \frac1{\imag qr}[B(\theta)\e^{\imag qr} - C(\theta) \e^{-\imag qr}] + \mathcal{O}\Bigl(\frac1{r^2}\Bigr),
\end{align}
where
\begin{align}
  B(\theta) &= 2\delta(1-\cos\theta) + \imag q f(q, \theta), \qquad
  C(\theta) = 2\delta(1+\cos\theta).
\end{align}
The optical theorem is obtained by the balance of the probability fluxes of the
outgoing and incoming waves:
\begin{align}
  0 &= \int r^2 \d\Omega \frac1{(qr)^2}\bigl[|B(\theta)|^2 - |C(\theta)|^2\bigr] \notag \\
    &= \int \d\Omega \frac1{q^2} \{2\Re[2 \imag q f(q, \theta) \delta(1-\cos\theta)] + q^2|f(q, \theta)|^2\} \notag \\
    &= -\frac{4\pi}q \Im f(q, \theta = 0) + \sigma_\mathrm{tot},
\end{align}
where the interference of the two terms in $B(\theta)$ picks up the forward
amplitude $f(q, \theta = 0)$.  The correlation function is given by the KP
formula:
\begin{align}
  C(q)
    &= \int_0^\infty r^2 \d{r} S(r) \int \d\Omega \frac1{(qr)^2}|B(\theta)\e^{\imag qr} - C(\theta)\e^{-\imag qr}|^2 \notag \\
    &= \int_0^\infty \d{r} S(r) \int \d\Omega \frac1{q^2} \{2|C(\theta)|^2 - 2\Re[B(\theta)C(\theta)^* \e^{2\imag qr}]\}.
\end{align}
By subtracting the plane-wave contribution, we obtain
\begin{align}
  C(q)
    &= 1 - \int_0^\infty \d{r} S(r) \int \d\Omega \frac1{q^2} 2\Re[2\imag q f(q, \theta) \delta(1+\cos\theta) \e^{2\imag qr}] \notag \\
    &= 1 + \frac{4\pi}q\int_0^\infty \d{r} S(r) \Im[f(q, \theta) \e^{2\imag qr}] \notag \\
    &= 1 + \frac{4\pi}q \Im[f(q, \theta) \hat S(-2\imag q)],
\end{align}
where we observe that the interference of $f(q, \theta)$ and $C(\theta)$ picks
up the backward amplitude $f(q, \theta = \pi)$.

\section{Numerical results}

\begin{figure}[htb]
  \centering
  \includegraphics[width = 0.85\textwidth]{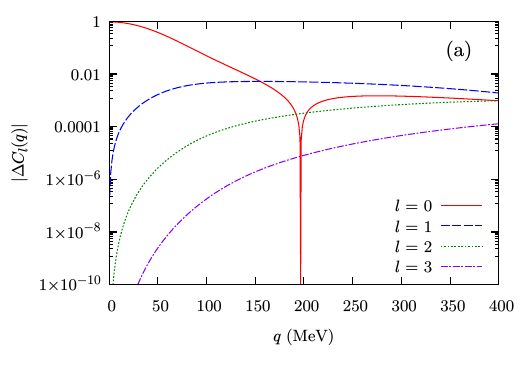}
  \includegraphics[width = 0.85\textwidth]{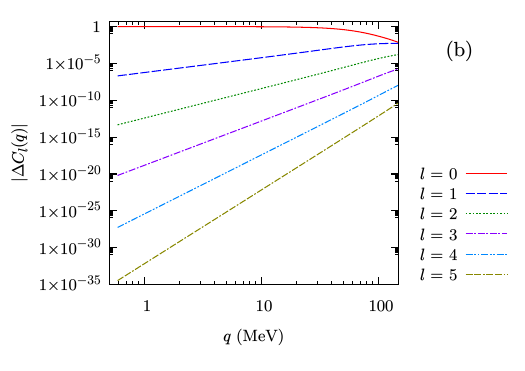}
  \caption{The magnitudes of the corrections~\eqref{eq:kpx.kpxd.delta} to the
  correlation function from each partial wave.  The reduced mass, the potential
  parameters, and the source size are chosen to be $\mu = 600~\text{MeV}$, $V_0
  = 50~\text{MeV}$, $b = 1~\text{fm}$, and $R = 2~\text{fm}$. The vertical axis
  shows the absolute value of the correction of each partial wave on a
  logarithmic scale.  The horizontal axis indicates the relative momentum $q$.
  Panel~(a) shows the behavior in a wide range of $q$, while panel~(b)
  shows the behavior at small $q$ with the logarithmic scale in the
  horizontal axis.}
  \label{fig:kpx1}
\end{figure}

For demonstration and examination of the formulae we obtained for the
higher partial waves, we consider a square potential well:
\begin{align}
  V(r) &= \begin{cases}
    -V_0, & (r < b), \\
    0, & (r > b),
  \end{cases}
\end{align}
with the depth $V_0$ and the width $b$ being the potential parameters.  We
assume the reduced mass $\mu = 600~\text{MeV}$ and the potential width $b =
1~\text{fm}$ in the following analyses.

\subsection{Spherical-source KP formula with higher partial waves}

We first consider the KP formula with the spherical source~\eqref{eq:kpx.kpxd}
and \eqref{eq:kpx.kpxd.delta}.  Figure~\ref{fig:kpx1} shows the magnitudes of the
corrections to the correlation function from different partial waves in the
case of the Gaussian source~\eqref{eq:ll.gaussian-source}.  The potential
parameters, $V_0 = 50~\text{MeV}$ and $b = 1~\text{fm}$, are chosen to be
weakly attractive so that no bound states appear.

At a lower relative momentum $q$ in Fig.~\ref{fig:kpx1}~(a), we see that the higher
orders of the partial waves are suppressed by orders of magnitude.  This is
because the wave function is given by $R_l(r), j_l(r) \approx q^l$ at the
leading order of $q$.  Therefore, the correction is expected to usually behave
in the limit $q\to 0$ as
\begin{align}
  \Delta C_l(q) \sim |R_l(r)|^2 - |j_l(qr)|^2 \sim q^{2l}.
\end{align}
This dependence is manifested in Fig.~\ref{fig:kpx1}~(b), where
the slope of each correction reads $2l$ in the log-log plot at small $q$.  This
means that the contributions from the higher partial waves are usually
negligible by orders of magnitude at small $q$.  This is also consistent with
the original paper of the LL formula~\cite{Lednicky:1981su}, which concluded
that the $s$-wave approximation works well for $q \le 50~\text{MeV}$ within the
1\% error.

However, the contributions from higher partial waves become comparable when the
relative momentum $q$ becomes larger.  In particular, when the lower-order
contributions of the partial waves crosses zero, the contribution of higher
partial waves can be larger than the lower orders.  For example, the
spike around $q\simeq200~\text{MeV}$ in the $l=0$ case on
Fig.~\ref{fig:kpx1}~(a) indicates that the $s$-wave correction $\Delta C_0(q)$
crosses zero.  In this case, the $p$ wave ($l=1$) has a larger contribution at
larger $q$.

\begin{figure}[htb]
  \centering
  \includegraphics[width = 0.85\textwidth]{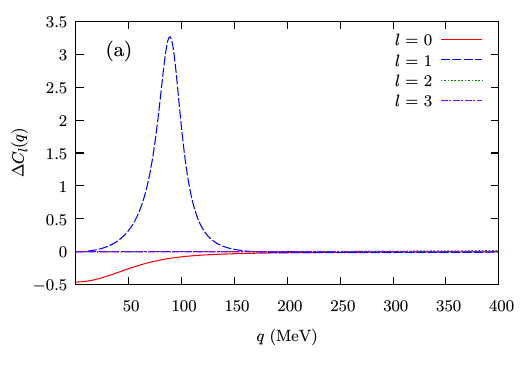}
  \includegraphics[width = 0.85\textwidth]{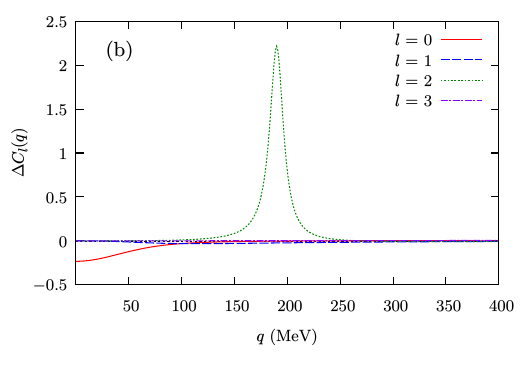}
  \caption{The same as Fig.~\ref{fig:kpx2} but with deeper potentials: $V_0 =
  300$ and $600~\text{MeV}$ for panels~(a) and~(b), respectively.}
  \label{fig:kpx2}
\end{figure}

It should be noted that the detailed behavior largely depends on the potential
and its parameters.  In particular, when there are resonances of higher partial
waves near the real axis in the $q$ plane, the peaks associated with the
resonances appear in the correlation function.  Figure~\ref{fig:kpx2} shows the
correlation function with deeper square potential well.  Panels~(a) and~(b)
show the results with $V_0 = 300$ and $600~\text{MeV}$, respectively.
Unlike the shallow case of $V_0 = 50~\text{MeV}$ in Fig.~\eqref{fig:kpx1}, the
red lines ($l=0$) are negative in both panels because of an $s$-wave bound
state.  We find significant peaks of the correlation function corrections by
the $p$ wave ($l=1$) and the $d$ wave ($l=2$) at $q \simeq 90~\text{MeV}$ on
panel~(a) and $q \simeq 190~\text{MeV}$ on panel~(b), respectively,
and the magnitude of the peaks are significantly larger than the magnitude of
the $s$-wave ($l=0$) correction.  In the case of $V_0 = 300~\text{MeV}$, the
$p$-wave contribution has almost the same magnitude as the $s$-wave
contribution even at a low momentum $q \simeq 50~\text{MeV}$\@.  This tells us
that the corrections of the higher partial waves can be more significant than
the $s$-wave contribution, depending on the details of the potential.

\subsection{Generalized LL formula for higher partial waves}

\begin{figure}[htb]
  \centering
  \includegraphics[width = 0.95\textwidth]{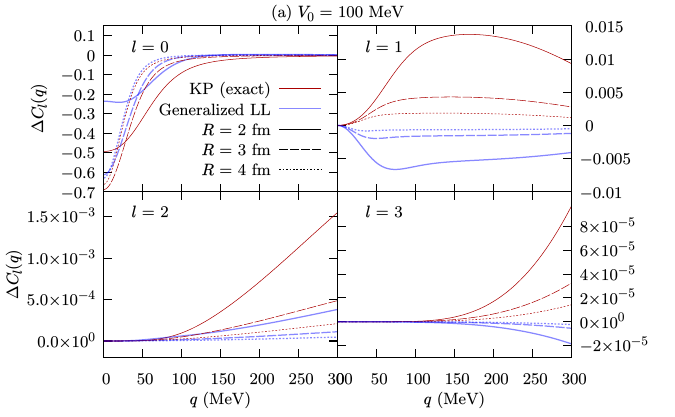}
  \includegraphics[width = 0.95\textwidth]{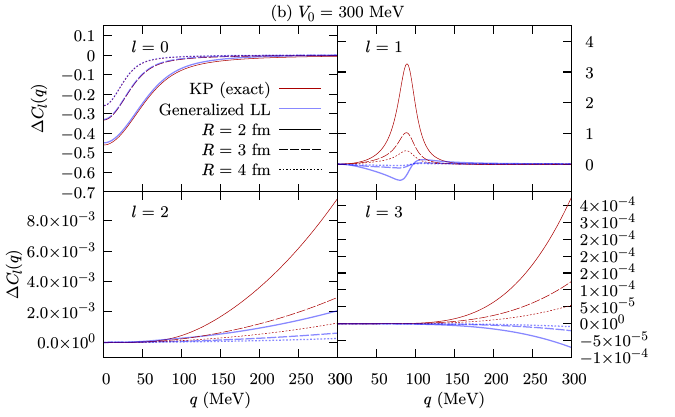}
  \caption{Comparison of the generalized LL formula to the corresponding
  correction in the KP formula with the direct integration.  Panels (a) and (b)
  show the results with $V_0 = 100$ and $300~\text{MeV}$, respectively.
  In each panel, the results for different angular momenta $l = 0, 1, 2, 3$ are
  shown from the top left to the bottom right.  The thin red and bold blue
  lines correspond to the KP and the generalized LL formulae, respectively.
  The solid, dashed, and dotted lines correspond to different source sizes $R =
  2$, $3$, and $4~\text{fm}$.  The other parameters are the same as in
  Fig.~\ref{fig:kpx1}.}
  \label{fig:llx1}
\end{figure}

We next examine the generalized LL formula~\eqref{eq:llx}.  While the partial
wave contributions of the KP formula~\eqref{eq:kpx.kpxd.delta} are exact under
the spherical source, the generalized LL
formula~\eqref{eq:llx} uses an assumption that the wave function in the entire
spatial region may be replaced by the $\mathcal{O}(1/r)$ asymptotic form
neglecting the centrifugal corrections.  This assumption appears to be highly
nontrivial to justify in the case of higher partial waves ($l \ge 1$).  Here, the
generalized LL formula shall be compared to the exact spherical-source KP
formula to examine whether it gives a reasonable approximation to the
spherical-source KP formula.

Figure~\ref{fig:llx1} compares the generalized LL formula to the partial wave
contribution of the KP formula.  Panels (a) and (b) correspond to different
potential depths, $V_0 = 100$ and $300~\text{fm}$, respectively.  We can first
confirm that the generalized LL and KP formulae for the $s$-wave case ($l=0$),
which are nothing but the traditional LL and KP formulae, are in good
agreement with each other.  It should be noted that the generalized LL formula
does not include the effective range correction, which was present in the
original LL formula, so the slight differences may be partly explained by
the effective range correction.

\begin{figure}[h]
  \centering
  \includegraphics[width = 0.80\textwidth]{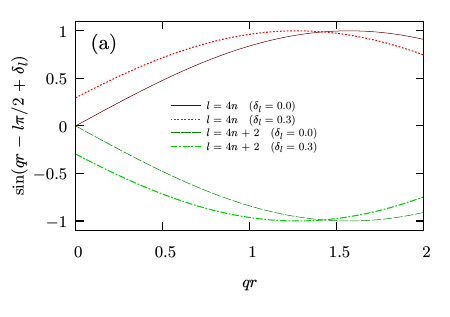}
  \includegraphics[width = 0.80\textwidth]{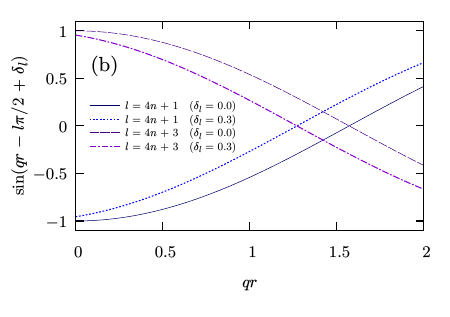}
  \caption{Phase shift effects on the asymptotic partial wave functions.
  Panels~(a) and~(b) show the cases of odd and even $l$'s, respectively.}
  \label{fig:sin}
\end{figure}

In the cases of higher partial waves ($l \ge 1$), however, the generalized LL
formula fails to reproduce the KP formula even qualitatively.  In particular,
the sign of the corrections are opposite for the odd $l$ due to the $(-1)^l$
factor in Eq.~\eqref{eq:llx}.  This factor $(-1)^l$ comes from the phase
$-l\pi/2$ in the asymptotic spherical Bessel function $j_l(qr) \approx \sin(qr -
l\pi/2)/(qr)$, ($r\to\infty$).  The correction to the correlation function
reflects the change of $|R^\mathrm{III}_l(r)|^2 \sim
|\sin(qr-l\pi/2+\delta_l)|/(qr)^2$ near the origin, where the source function
$S(r)$ has a support, by a finite $\delta_l$.  Figure~\ref{fig:sin} depicts the
effect of the phase shift on the asymptotic wave function of each angular
momentum in a modestly attractive case ($0 < \delta_l < \pi/2$).  We notice that the
magnitude of the wave function increases near the origin for even $l$, while it
decreases for odd $l$.  However, this behavior is unphysical.  First, the wave
function $R^\mathrm{III}_l(r) = \sin(qr-l\pi/2+\delta_l)/(qr)$ diverges at the
origin (except for very special values of the phase shift, $\delta_l=l\pi/2 +
n\pi$, $n\in\mathbb{Z}$).  This problem is already present with the assumption
behind the traditional LL formula.  More physically, the wave function near the
origin should be suppressed by the centrifugal force as $R_l(r) \approx r^l$,
($r\to0$), which is not taken into account in the wave function of the LL
formula.

This means that the generalized LL formula~\eqref{eq:llx} gives the
contribution of only a part of the wave function of the order
$\mathcal{O}(1/r)$ and misses an important part.  Although it gives an insight
into the structure of the LL formula as discussed in Sec.~\ref{sec:llopt}, it is
not practically useful for higher partial waves ($l \ge 1$) at all.  Even for the
traditional case for the $s$ wave ($l = 0$), the asymptotic wave function used
in the formula appears to have an unphysical behavior near the origin.  A
typical usage of the LL formula is to relate the correlation function to the
parameters of the effective range expansion, where the asymptotic wave function
can be regarded as merely one model of the wave function parametrized by the
phase shift or the parameters of the effective range expansion.  In this sense,
we can choose another more physical wave function as a reference model for
an improved LL formula.

\section{Summary}

Femtoscopy in high-energy collisions gives a novel and complementary approach
to the hadron interaction in combination with traditional scattering
experiments.  The understanding of the assumptions and applicability of
historical formulae used in femtoscopy is of great interest for future
analyses for the precise determination of the hadron interactions.  Important
formulae widely used in the analyses include the spherical-source KP
formula~\eqref{eq:kp.spherical}~\cite{Morita:2014kza} and the LL formula~\eqref{eq:ll.traditional}~\cite{Lednicky:1981su},
which assume that the two particles interact through only the $s$ wave ($l=0$).
However, it is nontrivial whether the higher partial waves ($l \ge 1$) are
negligible at all.  In this study, we considered the contributions of the
higher partial waves by extending the spherical-source KP
formula~\eqref{eq:kp.spherical} and the LL formula~\eqref{eq:ll.traditional}
assuming the KP formula~\eqref{eq:kpx.kp}.

We obtained the spherical-source KP formula with the contributions from all
partial waves in Eqs.~\eqref{eq:kpx.kpxd} and \eqref{eq:kpx.kpxd.delta}.  They
are summarized in the following form:
\begin{align}
  C(q)
  &= 1 + \sum_{l=0}^\infty (2l+1) \int \d{r}\, 4\pi r^2 S(r) \bigl[|R_l(r)|^2 - |j_l(qr)|^2\bigr],
\end{align}
where the correlation function gets the contributions of partial waves as the
sum of the contribution from each partial wave.

We also obtained the generalized version of the LL formula~\eqref{eq:llx}
(except for the effective range correction) by approximating the wave function
with the $\mathcal{O}(1/r)$ asymptotic form for the entire spatial region.
Furthermore, the sum for the angular momentum $l$ in Eq.~\eqref{eq:llx} can be
explicitly performed to obtain a more compact
representation~\eqref{eq:llopt.llg}.  These results can be summarized in the
following formula:
\begin{align}
  C(q)
    &= 1+ \frac{4\pi}q \Im [f(q, \theta = \pi) \hat S(-2\imag q)] \notag \\
    &= 1+ \frac{4\pi}q\sum_{l=0}^\infty (2l+1)(-1)^l \Im [f_l(q) \hat S(-2\imag q)],
\end{align}
where $\hat S(-2\imag q)$ is the Fourier--Laplace transform of the relative source
$S(r)$.  This reproduces the traditional LL formula~\eqref{eq:ll.traditional}
for the $s$-wave case, where two of the three correction terms come from $\Im
f_0(q) \Re \hat S(-2\imag q)$ (except for the effective range correction), and the
rest comes from $\Re f_0(q) \Im \hat S(-2\imag q)$.  We also point out that the
correction to the correlation function can be written by the backward
scattering amplitude $f(q, \theta = \pi)$, and the structure is similar to the
optical theorem.  We have shown that both the optical theorem and the correlation
function with the LL formula stem from an asymptotic structure of the plane
wave written by the forward and backward delta functions.  In particular, the
nontrivial parts in both cases come from interference between the scattering
amplitude $f(q, \theta)$ and the delta functions in the plane wave.

The corrections by the higher partial waves based on the KP
formula~\eqref{eq:kpx.kpxd.delta} were investigated for the square potential
well by changing the depth $V_0$.  In normal cases without resonances, the
contribution of higher partial waves to the correlation function is suppressed
by orders of magnitude at low momentum $q$, while it can be dominant at high
$q$ depending on the potential.  With resonances in higher partial waves ($l >
0$), the contribution of the higher partial wave can be more significant than
the $s$-wave ($l=0$) contribution.  The generalized LL formula~\eqref{eq:llx}
was also examined in comparison with the KP formula but turned out to fail for
higher partial waves ($l \ge 1$).  This is due to an unphysical behavior of the
asymptotic wave function applied to the interacting region $r \approx 0$.  This
suggests the necessity of an alternative wave-function form for an improved LL
formula, which will be discussed in our forthcoming paper.

\section*{Acknowledgement}
The authors thank Asanosuke Jinno and Yuki Kamiya for useful discussions.  The
work has been supported by JSPS KAKENHI Grant Numbers JP23H05439 (K.M. and
T.H.), JP23K13102 (K.M.), and JP22K03637 (T.H.).

\biboptions{numbers,sort&compress}
\bibliographystyle{elsarticle-num}
\bibliography{proc_higher}
\end{document}